\documentclass[pra,twocolumn,showpacs]{revtex4-1}
\usepackage{amsmath,amssymb,graphicx}

\newcommand{\Tr}{\textrm{Tr}}
\newcommand{\ket}[2][]{{|#2\rangle_{#1}}}
\newcommand{\bra}[2][]{{}_{#1}\langle #2|}

\newcommand{\proj}[2][]{\ket{#2}_{#1}\bra{#2}}

\begin{document}

\title{Linear optics schemes for entanglement distribution\\with realistic single-photon sources}

\author{Miko{\l}aj Lasota}
\email{Corresponding author. E-mail: miklas@fizyka.umk.pl}
\affiliation{Faculty of Physics, Astronomy and Applied Informatics, Nicolaus Copernicus University, Grudziadzka~5, 87-100~Toru\'{n}, Poland}

\author{Czes{\l}aw Radzewicz}
\author{Konrad Banaszek}
\affiliation{Faculty of Physics, University of Warsaw, Ho\.{z}a~69, 00-681~Warsaw, Poland}

\author{Rob Thew}
\affiliation{Group of Applied Physics, University of Geneva, 1211 Geneva 4, Switzerland}

\begin{abstract}
We study the operation of linear optics schemes for entanglement distribution based on nonlocal photon subtraction when input states, produced by imperfect single-photon sources, exhibit both vacuum and multiphoton contributions. Two models for realistic photon statistics with radically different properties of the multiphoton ``tail'' are considered. The first model assumes occasional emission of double photons and linear attenuation, while the second one is motivated by heralded sources utilizing spontaneous parametric down-conversion. We find conditions for the photon statistics that guarantee generation of entanglement in the relevant qubit subspaces and compare it with classicality criteria. We also quantify the amount of entanglement that can be produced with imperfect single-photon sources, optimized over setup parameters, using as a measure entanglement of formation. Finally, we discuss verification of the generated entanglement by testing Bell's inequalities. The analysis is carried out for two schemes. The first one is the well-established one-photon scheme, which produces a photon in a delocalized superposition state between two nodes, each of them fed with one single photon at the input. As the second scheme, we introduce and analyze a linear-optics analog of the robust scheme based on interfering two Stokes photons emitted by atomic ensembles, which does not require phase stability between the nodes.
\end{abstract}


\maketitle

\section{Introduction}
\label{Sec:Introduction}

One of the grand challenges in emerging quantum technologies is the distribution of entanglement over long distances, which would significantly enhance the feasible range of quantum key distribution \cite{GisiRiboRMP2002,ScarBechRMP2009} and other quantum communication protocols. Loss and other decoherence mechanisms, inevitably affecting long-haul transmission of quantum systems, e.g., light in optical fibers, dramatically attenuate available nonclassical correlations with an increasing distance. Presently, the most promising solution to this problem is an architecture based on a sequence of quantum repeaters placed at regular intervals over the distance to be covered \cite{BriDurPRL1998}. First, entanglement is generated between quantum memories located at adjacent nodes, which can be done more efficiently owing to shorter separation. In the second stage, entanglement swapping operations \cite{BennBrasPRL1993,ZukoZeilPRL1993} performed on quantum memories at individual nodes create long-distance entanglement between the end stations.

A natural choice for the physical implementation of quantum repeaters are atomic ensembles \cite{LukinRMP2003} or solid-state systems for storing quantum superpositions, with optical interconnects to facilitate transfers of quantum states \cite{SangSimonRMP2011}.
An essential ingredient in quantum repeater architectures is a scheme to distribute entanglement between adjacent nodes. Proposed designs are based on Raman scattering in atomic ensembles \cite{DuanLukinNAT2001,JiangTay07,ZhaoChen07,ChenZhao07,SangSimon08}, photon pair sources combined with quantum memories \cite{SimondeRied07}, as well as on a combination of multiple single-photon sources, memories, linear optics, and conditional photodetection \cite{SangSimon07}. In the last case, the underlying idea is to perform nonlocal photon subtraction \cite{DaknaAnhut97,KimPark05,Kim08}, which creates, in a heralded, loss-tolerant way, entangled states of excitations stored in quantum memories. The single photons at the input are a nonclassical resource which allows for the generation of entanglement.

The purpose of this paper is to analyze imperfections of single-photon sources that can be tolerated in the operation of linear-optics schemes for entanglement distribution based on nonlocal photon subtraction. Our analysis is based on two examples. The first one is the original proposal by Sangouard {\em et al.} \cite{SangSimon07} to prepare a superposition of two distant quantum memories sharing a single excitation. As the second example, we introduce a linear optics version of the scheme based on two-photon interference of Stokes photons emitted from atomic ensembles \cite{ChenZhao07,ZhaoChen07} and study its robustness against source imperfections.

Linear optics schemes for entanglement distribution are sensitive to both vacuum and multiphoton contributions in the input photon statistics. We will consider here two models of the photon number distribution for sources used to generate entanglement. The first model is a statistical mixture of up to two photons, that includes both non-ideal photon preparation and the possibility of double photon emission. The second model is motivated by heralded single-photon sources based on spontaneous parametric down-conversion \cite{CastellettoScholten}. Their typically low success rates could be improved in principle through the construction of multiplexed arrays \cite{MigdallBranningPRA2002,ShapiroWongOL2007,CollinsXiongNCOMM2013}. The down-conversion model of photon statistics exhibits a relatively long multiphoton ``tail'' vanishing more slowly than the thermal distribution with an increasing photon number. Considering two models with radically different properties in the multiphoton sector will allow us to assess whether the specifics of the multiphoton contribution may have a noticeable impact on the entanglement distribution scheme. The actual statistics of sources that are currently being developed \cite{ShieldsNPHOT2007,EisamanFanRSI2011} can be expected to interpolate between the two extreme models studied here.

To characterize the effects of imperfections, we quantify the entanglement generated in the relevant qubit subspaces with the help of entanglement of formation \cite{HillWoot97,Woot98}. This measure can be computed for a pair of two-level systems in a straightforward manner, providing an upper bound for distillable entanglement \cite{HoroHoroRMP2009}. We also give simple threshold criteria for the photon statistics necessary for entanglement generation at all. It is interesting to discuss these criteria in the context of nonclassicality of input light: clearly, if the sources exhibited Poissonian statistics, with optional classical excess noise, no entanglement generation would be possible, as a setup based on such sources, linear optics, and standard photodetectors would admit an entirely semiclassical description \cite{MandelJOSA1977}. We find that the non-classicality condition and the criteria for entanglement generation are not equivalent.

As the last aspect, we discuss the possibility to verify successful entanglement generation by demonstrating a violation of Bell's inequalities \cite{Bell87}. For the single-photon superposition generated between two quantum memories, we use the Clauser-Horne inequality \cite{ClauHorne74} applied to unbalanced homodyning measurements of phase-space quasiprobability distribution functions \cite{BanaWodPRL1999}. In this approach, noncommuting observables are realized with the help of phase-space displacements introduced by superposing the signal field with a coherent reference on an unbalanced beam splitter \cite{WallenVog96,BanaWod96}. In the second example, when the quantum memory at each mode contains an excitation prepared in a superposition of two modes, we can use the standard Clauser-Horne-Shimony-Holt (CHSH) inequality \cite{CHSH69}, although care needs to be taken to account correctly for multiphoton terms.

This paper is organized as follows. First, in Sec.~\ref{Sec:Subtracting} we present a general theoretical description of linear-optics schemes for nonlocal photon subtraction. Simplified formulas in the high-loss regime that are most relevant in practical scenarios are derived. Section III introduces two models of photon statistics that will be used to study the effects of source imperfections. The generic one-photon scheme is described in Sec.~\ref{Sec:TwoMemories}, where a threshold criterion for the photon statistics to warrant entanglement generation is also derived. A quantitative characterization of entanglement produced in this scheme and requirements to violate phase-space Bell's inequalities are presented in Sec.~\ref{Sec:OnePhotonEntanglement}. A linear optics two-photon scheme robust against phase instabilities is introduced in Sec.~\ref{Sec:FourMemories}. Its properties are discussed quantitatively in Sec.~\ref{TwoPhotonEntanglement}.
Finally, Sec.~\ref{Sec:Conclusions} concludes the work.

\section{Nonlocal photon subtraction}
\label{Sec:Subtracting}

\begin{figure}
\centering
\includegraphics[width=3.3in]{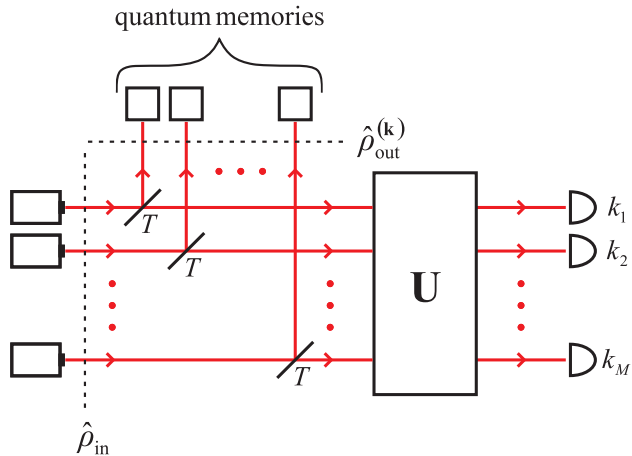}
\caption{(Color online) A general scheme for nonlocal photon subtraction that includes schemes for entanglement distribution studied in this work. The light emitted by an array of single-photon sources located on the left of the diagram is sent to beam splitters with identical power transmission coefficients $T$. Reflected beams are mapped onto local quantum memories, while transmitted beams travel to an intermediary site, where they enter a linear optics circuit described by a unitary matrix ${\mathbf U}$ with outputs monitored by heralding detectors that generate counts $k_1, \ldots , k_M$.}
\label{fig:figure1}
\end{figure}

A number of linear optics schemes for entanglement distribution via nonlocal photon subtraction can be described using a general setup depicted in Fig.~\ref{fig:figure1}(a). An array of $M$ sources emits single photons in well-defined modes represented by respective annihilation operators $\hat{a}_i$, $i=1,\ldots, M$. The sources are divided into two groups located at adjacent nodes. At each node, the photons are directed to beam splitters with identical power transmissions $T$. The transmitted beams are sent to an intermediary site, where they are combined using a linear optical circuit characterized by a certain unitary $M \times M$ matrix $\mathbf{U}$, whose outputs are monitored by an array of single-photon detectors. We assume here that losses affecting all the modes between the beam splitters and the intermediary site, as well as the efficiencies of the detectors, are uniform. The beams reflected off the beam splitters placed after single-photon sources are mapped locally onto quantum memories present at the nodes. Light stored in quantum memories is retained for further processing only if the detectors at the intermediary site produce certain sequences of clicks, heralding that entanglement between memories has been successfully generated. In subsequent stages, entanglement swapping operations can be used to extend the range of entanglement to more distant nodes.

Let us denote by $\hat{b}_i$, $i=1,\ldots, M$, annihilation operators of modes after the transformation $\mathbf{U}$. Their relation to the input operators $\hat{a}_i$ can be written compactly as
\begin{equation}
\begin{pmatrix} \hat{b}_1 \\ \hat{b}_2 \\ \vdots \\ \hat{b}_M \end{pmatrix} = \mathbf{U}
\begin{pmatrix} \hat{a}_1 \\ \hat{a}_2 \\ \vdots \\ \hat{a}_M \end{pmatrix}.
\end{equation}
Suppose that the detectors produce a specific sequence of clicks described by a vector ${\mathbf k} = (k_1, k_2, \ldots, k_M)$. The transformation between the input state $\hat{\varrho}_{\text{in}}$ and the unnormalized conditional output state
${\hat{\varrho}^{({\mathbf k})}}_{\text{out}}$ generated at the reflected ports of the beam splitters is given by a map:
\begin{equation}
{\hat{\varrho}^{({\mathbf k})}}_{\text{out}} = \sum_{{\mathbf n}} p({\bf k}|{\bf n}) \hat{L}_{\mathbf n} \hat{\varrho}_{\text{in}} \hat{L}_{\mathbf n}^\dagger. \label{Eq:rho(k)out}
\end{equation}
Here $\hat{L}_{\mathbf n}$ are Kraus operators corresponding to the subtraction of $n_i$ photons from the mode $\hat{b}_i$, $i=1,\ldots, M$
\cite{ChuangLeungPRA1997,WasiBanaPRA2007},
\begin{equation}
\hat{L}_{\bf n} = \bigotimes_{i=1}^{M} \sqrt{\frac{T^{n_i}}{n_i!}} (\sqrt{1-T})^{\hat{b}^\dagger_i \hat{b}_i} \hat{b}_i^{n_i}
\label{Eq:Kraus}
\end{equation}
and we have denoted $\mathbf{n} = (n_1, n_2, \ldots, n_M)$. Further, $p({\bf k}|{\bf n})$ is the conditional probability of producing a click sequence ${\bf k}$ on the detectors, provided that $n_1,  \ldots, n_M$ photons have been subtracted from the modes $\hat{b}_1 , \ldots, \hat{b}_M$. Its form depends on the specific detection scheme used in the setup.

Let us denote by $\zeta$ the detection efficiency, assumed to be identical for all the detectors. As the transformation $\mathbf{U}$ is linear, the parameter $\zeta$ can also account for losses between the nodes and the intermediary site, following the assumption that losses are identical for all the modes involved. If photon-number-resolving detectors  are used at the intermediary site, the conditional probability $p({\bf k}|{\bf n})$ is given by a multimode binomial distribution
\begin{equation}
p({\bf k}|{\bf n}) = \prod_{i=1}^{M} { n_i \choose k_i } \zeta^{k_i} (1-\zeta)^{n_i - k_i}
\end{equation}
where we have used convention ${n \choose k } = 0$ for $n < k$. Throughout this paper we will focus our attention on the second relevant case, when binary detectors are used at the intermediary site. Then each $k_i$ can take only two values, $0$ or $1$, depending on whether the respective detector has registered no photons or at least one, and the conditional probability distribution $p({\bf k}|{\bf n})$ takes the form
\begin{equation}
\label{Eq:CondProbBinary}
p({\bf k}|{\bf n}) = \prod_{i=1}^{M} [k_i + (1-2 k_i)(1-\zeta)^{n_i}].
\end{equation}
In a realistic scenario losses between the nodes and the intermediary site are significant and consequently $\zeta \ll 1$. In this regime, in Eq.~(\ref{Eq:CondProbBinary}) we can approximate $(1-\zeta)^{n_i} \approx 1-\zeta n_i$, provided that the photon numbers at the setup input are not too high. Taking the leading order in $\zeta$ of factors appearing under the product in Eq.~(\ref{Eq:CondProbBinary}), which is $1$ for $k_i=0$ and $\zeta n_i$ for $k_i=1$, gives a simplified expression for the conditional probability distribution for $\zeta \ll 1$,
\begin{equation}
\label{Eq:CondProbBinaryLowEta}
p({\bf k}|{\bf n}) = \zeta^{K} \prod_{i=1}^{M} (1-k_i + k_i n_i)
\end{equation}
where $K= \sum_{i=1}^{M} k_i$ is the total number of clicks produced by the binary detectors. The above formula gives an explicit power scaling in the efficiency $\zeta$.

The principle of operation for entanglement distribution based on nonlocal photon subtraction is typically discussed in the limit of very high reflection of produced photons to quantum memories, corresponding to $T \rightarrow 0$. In this case, we can approximate
$(\sqrt{1-T})^{\hat{b}^\dagger_i \hat{b}_i} \approx 1$ in Eq.~(\ref{Eq:Kraus}), assuming that contributions from higher photon numbers at the input are sufficiently small \cite{SekatskiJPB2012}. Then the conditional state is given by
\begin{multline}
{\hat{\varrho}^{({\mathbf k})}}_{\text{out}} = \sum_{{\mathbf n}} p({\bf k}|{\bf n})
\frac{T^{n_1 + \cdots + n_M}}{n_1 ! \cdots n_M!} \\
\times \hat{b}_1^{n_1} \cdots \hat{b}_M^{n_M} \hat{\varrho}_{\text{in}} (\hat{b}_1^\dagger)^{n_1} \cdots (\hat{b}_M^\dagger)^{n_M}.
\label{eq:KrausForSmallT}
\end{multline}
When $T\rightarrow 0$, the summation over $\mathbf{n}$ in the above formula can be truncated to the lowest value of the total photon number
$n_1 + \cdots + n_M$ that gives a nonzero contribution to a click sequence $\mathbf{k}$. This yields a simple power scaling in $T$.

In the following, we will consider the physical regime when $\zeta \ll 1$, as otherwise formulas become overly complicated. We will use the assumption $T \ll 1$ mainly for illustrative purposes. We will carry out numerical calculations for arbitrary $T$ using the general expression given in Eq.~(\ref{Eq:Kraus}) in order to optimize the amount of generated entanglement.


\section{Imperfect photon sources}
\label{Sec:Sources}

We will consider in this paper two models for photon number distribution produced by an imperfect source. The most elementary model, which includes both non-unit preparation efficiency and multiphoton effects, is a statistical mixture of up to two photons, $p_0 \proj{0} + p_1 \proj{1} + p_2 \proj{2}$, where the three probabilities $p_0, p_1$, and $p_2$, satisfy the normalization constraint $p_0 + p_1 + p_2 = 1$. The main advantage of this elementary model is computational simplicity.

It will be convenient to introduce a physically motivated parametrization for the photon number distribution in the above model. Namely, we will assume that the statistics are generated by a source producing a single photon with a probability $1-\epsilon$, while a double photon emission occurs with a probability $\epsilon$, and that the output is subject to linear losses characterized by a power transmission coefficient $\eta$. The explicit expressions for photon number probabilities in this double-emission model are
\begin{subequations}
\label{Eq:UptoTwoPhotons}
\begin{align}
p_0 & = (1-\eta)(1-\eta\epsilon), \\
p_1 & = \eta + \eta(1-2\eta)\epsilon, \\
\label{Eq:UptoTwoPhotonsp2}
p_2 & = \eta^2 \epsilon.
\end{align}
\end{subequations}

The second model is motivated by heralded sources based on nondegenerate spontaneous parametric down-conversion \cite{CastellettoScholten}, when photons within produced pairs can be separated by polarization, frequency, or the emission direction. If the process involves only one pair of field modes for the signal and the idler beams, the probability of generating simultaneously $n$ pairs scales as $r^n$, where $r$ depends on the strength of the nonlinear process.

Suppose that the idler beam is monitored by a binary heralding detector with a very low efficiency. In this case, the conditional probability that the detector clicks scales linearly with the number of incident idler photons, which is a special case of Eq.~(\ref{Eq:CondProbBinaryLowEta}) for $M=1$. Consequently, the probability of producing $n$ photons in the heralded signal beam is given, after normalization, by $(1-r)^2 n r^{n-1}$. This statistics features a ``tail'' for $n > 2$, which vanishes more slowly than Poissonian and thermal distributions with increasing $n$. If the signal beam experiences losses characterized by a power transmission coefficient $\eta$, the photon statistics of the source is given by
\begin{align}
p_m & = (1-r)^2 \sum_{n=m}^{\infty} { n \choose m} \eta^m (1-\eta)^{n-m} n r^{n-1} \nonumber \\
 & =\frac{(1-r)^2 r^{m-1} \eta^m [m+(1-\eta) r]}{[1-(1-\eta)r]^{m+2}}.
\label{eq:probdistribution}
\end{align}
This distribution describes the second, down-conversion model used in our analysis.

It is easy to check that when $r \ll 1$, expanding Eq.~(\ref{eq:probdistribution}) for $m=0,1,2$ up to the linear term in $r$ yields Eq.~(\ref{Eq:UptoTwoPhotons}) with $r=\epsilon/2$. Furthermore, we verified numerically that for $\eta \ge 0.834$ the overall multiphoton probability $\sum_{m=2}^{\infty} p_m$ for the down-conversion model is slightly smaller than the two-photon probability given by Eq.~(\ref{Eq:UptoTwoPhotonsp2}), with $\epsilon = 2r < 0.5$. Consequently, we can compare the performance of entanglement distribution schemes for the two models of the photon statistics with their parameters, identified as $r=\epsilon/2$, to estimate the effects of the actual shape of the multiphoton ``tail'' produced by an imperfect photon source.

The parameter $\eta$, common to both the models of the photon statistics, can be used to account for attenuation along the optical path from a photon source to a respective quantum memory, as well as linear losses associated with a mapping onto the memory and its subsequent readout. This parameter critically affects the quality of the generated entanglement. In contrast, the primary effect of losses along the path from the transmitted output ports of beam-splitters $T$ to detectors at the intermediary site, characterized by the transmission $\zeta$ introduced in Sec.~\ref{Sec:Subtracting}, is the reduction of the rate of heralding events.

\section{One-photon scheme}
\label{Sec:TwoMemories}

\begin{figure}
\centering
\includegraphics[width=3in]{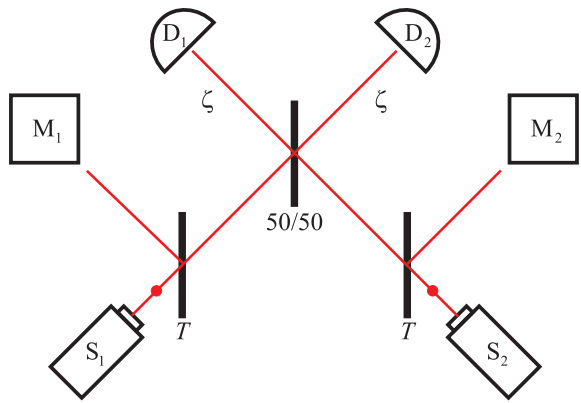}
\caption{(Color online) A scheme for producing a single photon in a delocalized superposition state between two quantum memories, originally proposed in Ref.~\cite{SangSimon07}. $S_1, S_2$, single-photon sources; $M_1, M_2$, quantum memories; $D_1, D_2$, heralding detectors. }
\label{fig:figure2}
\end{figure}

The simplest scheme for generating entanglement between two nodes, shown in Fig.~\ref{fig:figure2},  has been proposed by Sangouard \emph{et al.} in \cite{SangSimon07}. Each node has one single-photon source and a memory. We will label the corresponding modes  with indices $a_1$ and $a_2$ for the two nodes. The beams sent to the intermediary site are interfered on a balanced $50/50$ beam splitter described by a unitary transformation
\begin{equation}
\mathbf{U} = \frac{1}{\sqrt{2}} \begin{pmatrix} 1 & 1 \\ 1 & -1 \end{pmatrix}.
\end{equation}
We will restrict our attention here to the regime $\zeta \ll 1$ and binary detectors monitoring the two output ports of the beam splitter. A calculation based on Eqs.~(\ref{Eq:rho(k)out}) and (\ref{Eq:CondProbBinaryLowEta}) shows that if perfect single-photon sources are used in the setup with the initial state $\hat{\varrho}_{\text{in}} = \proj{1_{a_1} 1_{a_2}}$, a single detector count implies generation of one of two conditional states, denoted here with the superscript $\pm$:
\begin{equation}
\label{Eq:rhooutpmOnePhoton}
\hat{\varrho}_{\text{out}}^{(\pm)} = \zeta T[ (1-T) \proj{\psi_\pm} + T \proj{\text{vac}}]
\end{equation}
where the first component is a  maximally entangled state obtained by non-local photon subtraction,
\begin{align}
\label{Eq:psipm}
\ket{\psi_\pm}  & = \frac{1}{\sqrt{2}} (\hat{a}_1 \pm \hat{a}_2) \ket{1_{a_1} 1_{a_2}} \nonumber \\
 &= \frac{1}{\sqrt{2}} (\ket{0_{a_1} 1_{a_2}} \pm \ket{1_{a_1} 0_{a_2}})
\end{align}
with the sign of the superposition depending
on the detector that clicked, and $\ket{\text{vac}}$ denotes the vacuum state. In the limit $T\rightarrow 0$ the relative weight of the vacuum component with respect to $\ket{\psi_\pm}$ becomes negligible. This results in production of the ideal maximally entangled state, albeit with a diminishing overall success probability scaling as the product $\zeta T$.

If the photon sources are imperfect with their statistics described by identical photon number distributions $p_m$, the initial state of the modes $a_1$ and $a_2$ is given in the general case by a density matrix
\begin{equation}
\hat{\varrho}_{\text{in}} = \sum_{m_1, m_2=0}^{\infty} p_{m_1} p_{m_2} \proj[{a_1}]{m_1} \otimes \proj[{a_2}]{m_2}.
\label{eq:rhoin}
\end{equation}
Assuming that binary detectors are used at the intermediary site,
we will be interested in the conditional states $\hat{\varrho}^{({\mathbf k})}$ defined in Eq.~(\ref{Eq:rho(k)out}) for ${\mathbf k} = (1,0)$ and
${\mathbf k} = (0,1)$. These states are identical up to a trivial $\pi$ phase shift performed on one of the modes, which of course does not change the amount of the generated entanglement. Following the notation introduced in Eq.~(\ref{Eq:rhooutpmOnePhoton}), we will label them with a superscript $\pm$ in lieu of ${\mathbf k}$.

Let us first analyze the entanglement present in the qubit subspace spanned by the zero- and one-photon Fock state for each of the stored modes
when imperfect single-photon sources are used. This characteristic is contained in the projected memory state, denoted here with a prime,
\begin{equation}
\mbox{${\hat{\varrho}'}$}^{(\pm)}_{\text{out}}  = ( \hat{\Pi}_{a_1} \otimes \hat{\Pi}_{a_2}) \hat{\varrho}^{(\pm)}_{\text{out}}
( \hat{\Pi}_{a_1} \otimes \hat{\Pi}_{a_2})
\end{equation}
where the projection operators are
\begin{equation}
\hat{\Pi}_{\mu} = \proj[\mu]{0} + \proj[\mu]{1}, \quad \mu=a_1, a_2.
\label{eq:projection1ph}
\end{equation}
In the two-mode subspace spanned by $\ket{0_{a_1} 0_{a_2}}$, $\ket{0_{a_1} 1_{a_2}}$, $\ket{1_{a_1} 0_{a_2}}$, $\ket{1_{a_1} 1_{a_2}}$, the projected density matrix is given by
\begin{equation}
\mbox{${\hat{\varrho}'}$}^{(\pm)}_{\text{out}}
= \begin{pmatrix} \varrho_{00} & 0 & 0 & 0 \\ 0 & \varrho_{01} & c & 0 \\ 0 & c^\ast & \varrho_{10} & 0 \\ 0 & 0 & 0 & \varrho_{11}
\end{pmatrix},
\label{eq:rhoprim1ph}
\end{equation}
where the only non-zero elements in the limit $T\ll 1$ take the form
\begin{subequations}
\begin{align}
{\varrho}_{00} & = \zeta Tp_0p_1, \\
{\varrho}_{01} = {\varrho}_{10} & = \zeta T\left(p_0p_2+\frac{p_1^2}{2}\right), \\
\varrho_{11} & = 2 \zeta  T p_1p_2, \\
c & = \pm \frac{1}{2} \zeta T p_1^2.
\end{align}
\end{subequations}
The positive partial transposition (PPT) criterion \cite{Peres96,Horodeckis96} implies that the state $\mbox{${\hat{\varrho}'}$}^{(\pm)}_{\text{out}}$  is entangled if and only if
\begin{equation}
|c|^2>\varrho_{00}\varrho_{11}. \label{eq:PPT}
\end{equation}
This inequality translates into a simple condition for the photon statistics:
\begin{equation}
\label{Eq:EntanglementGenIneq}
p_1 ^2 > 8 p_0 p_2
\end{equation}
that ensures generation of entanglement in the limit $T \ll 1$.

It is interesting to compare the above inequality with properties of the photon number distribution for classical light states, i.e., coherent states and their statistical mixtures. In the latter case, the weight of an $n$-photon term can be expressed as
\begin{equation}
p_n = \left\langle \frac{{\cal I}^n}{n!} e^{-{\cal I}} \right\rangle,
\end{equation}
where ${\cal I}$ is the mean photon number of a coherent state and angular brackets $\langle \ldots \rangle$ denote a classical statistical average over ${\cal I}$. Using the Schwarz inequality $\langle XY \rangle \le \langle X^2 \rangle \langle Y^2 \rangle$ for $X = e^{-{\cal I}/2}$ and $Y = {\cal I}e^{-{\cal I}/2}$ yields
\begin{equation}
p_1^2 \le 2 p_0 p_2.
\label{Eq:CoherentMixtures}
\end{equation}
It is worth noting a gap between the above classicality condition and the parameter region characterized by Eq.~(\ref{Eq:EntanglementGenIneq}) for which photon sources are capable of producing entanglement in the one-photon scheme considered here.

\begin{figure*}
\centering
\includegraphics[width=14cm]{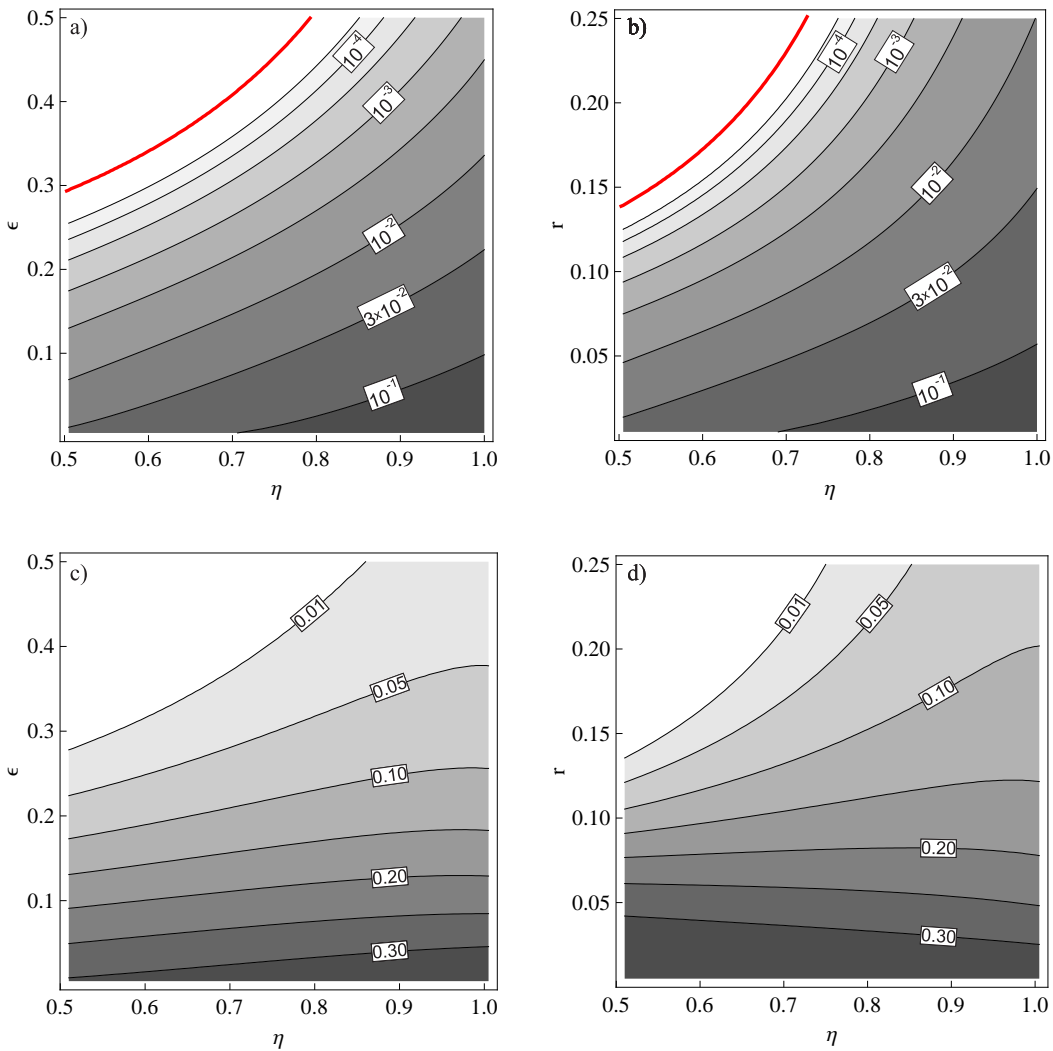}
\caption{(Color online) The effective entanglement of formation $E$ produced in the subspace spanned by zero- and one-photon Fock states for the input photon statistics  (a) described by a mixture of up to two photons defined in Eq.~(\ref{Eq:UptoTwoPhotons}) and (b) based on the process of heralded parametric down-conversion, according to Eq.~(\ref{eq:probdistribution}). Thick solid lines (red online) depict the criterion (\ref{Eq:EntanglementGenIneq}) that warrants entanglement generation in the limit $T\ll 1$. Beam-splitter transmissions $T$ maximizing the effective entanglement of formation for given parameters of the photon statistics are shown respectively in panels (c) and (d).}
\label{fig:figure3}
\end{figure*}

The condition for entanglement generation retains a relatively compact form for an arbitrary $T$ if the photon sources produce statistical mixtures of up to two photons. Using more general expressions for the elements of the density matrix $\mbox{${\hat{\varrho}'}$}^{(\pm)}_{\text{out}}$ in this case, the entanglement criterion given in Eq.~(\ref{eq:PPT}) can be generalized to
\begin{multline}
p_1^2\left(\frac{p_1^2}{4}-2p_0p_2\right) >  4 T^4 p_2^4 + 8 T^3 p_1 p_2^3  \\
 + 4 T^2 p_2^2 (p_1^2 +2 p_0 p_2) + 8 T p_0 p_1 p_2^2. \label{eq:PPTgeneral}
\end{multline}
It is seen that the right-hand side of the inequality is a fourth-order polynomial in $T$ with non-negative coefficients.
Consequently, the resulting condition on photon statistics becomes the most relaxed in the limit $T\rightarrow 0$, when it can be written simply as
Eq.~(\ref{Eq:EntanglementGenIneq}). We also verified by a direct calculation that if the conditional probability distribution $p({\bf k}|{\bf n})$ is taken in the general form given by Eq.~(\ref{Eq:CondProbBinary}) that is valid outside the regime $\zeta \ll 1$, then the right-hand side of Eq.~(\ref{eq:PPTgeneral}) also has the form of a polynomial in $T$ with non-negative coefficients. Consequently, the condition presented in Eq.~(\ref{Eq:EntanglementGenIneq}) is least restrictive over arbitrary values of $\zeta$ and $T$.
In Figs.~\ref{fig:figure3}(a) and \ref{fig:figure3}(b) we depict with thick solid lines (red online) the condition for entanglement generation given in Eq.~(\ref{Eq:EntanglementGenIneq}) in terms of parametrizations of the two models of photon statistics discussed in Sec.~\ref{Sec:Sources}. We will present a more quantitative characterization of the generated entanglement in the next section.

\section{One-photon entanglement}
\label{Sec:OnePhotonEntanglement}

Let us now discuss in more detail the amount of entanglement produced in the one-photon scheme and the possibility to verify its successful generation using Bell's inequalities.

\begin{figure*}
\centering
\includegraphics[width=14cm]{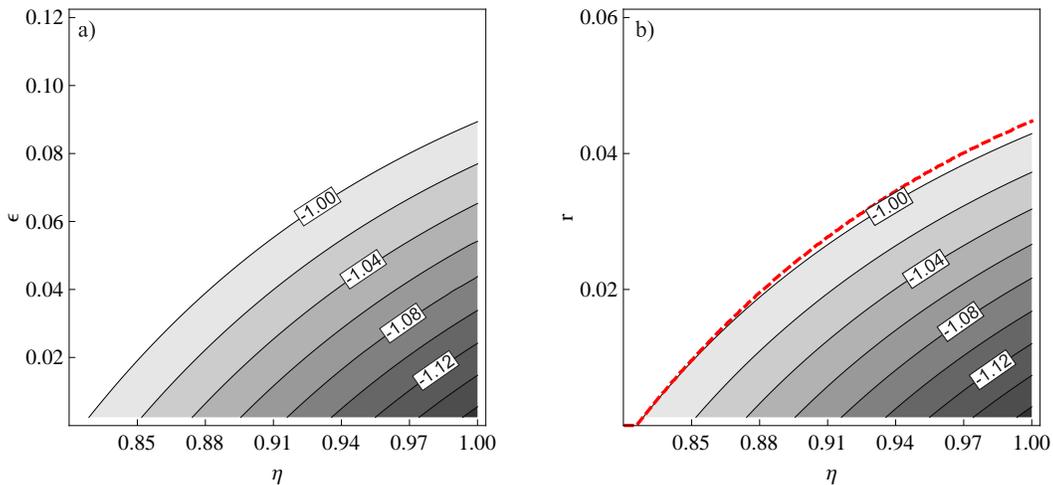}
\caption{The minimum value of the Clauser-Horne combination ${\cal CH}$ for coherent-state projections performed on the one-photon entangled state generated using the scheme presented in Fig.~\ref{fig:figure2}, assuming (a) the double-emission model for the photon statistics given in Eq.~(\ref{Eq:UptoTwoPhotons}) and (b) the down-conversion model derived in Eq.~(\ref{eq:probdistribution}). The dashed line in panel (b) indicates the contour ${\cal CH}= -1$ from panel (a). Note the different scales on the axes compared to Fig.~\ref{fig:figure3}.}
\label{fig:figure4}
\end{figure*}

As an entanglement measure, we choose the entanglement of formation $E_F$, which is defined as the number of maximally entangled states needed to prepare an ensemble of pure states representing a given mixed state, minimized over all such ensembles. For a normalized two-qubit state $\hat{\varrho}$ the entanglement of formation is given explicitly by \cite{HillWoot97,Woot98}
\begin{equation}
E_F(\hat\varrho)=H\left({\textstyle\frac{1}{2}} \bigl( 1 +\sqrt{1-[C(\hat\varrho)]^2}\bigr)\right),
\end{equation}
where $H(x)= -x \log_2 x - (1-x) \log_2 (1-x)$ denotes binary entropy and $C(\hat\varrho)$ is the so-called concurrence of the state $\hat\varrho$, which can be computed in a straightforward manner. The effective amount of entanglement which includes the nonunit probability of generating successfully the desired state is
$2 \Tr (\mbox{${\hat{\varrho}'}$}^{(\pm)}_{\text{out}}) E_F ( \mbox{${\hat{\varrho}'}$}^{(\pm)}_{\text{out}} /
\Tr (\mbox{${\hat{\varrho}'}$}^{(\pm)}_{\text{out}})) $, where the factor $2$ comes from two possible types of detection events denoted in Sec.~\ref{Sec:TwoMemories} as $\pm$. In the limit $\zeta \ll 1$ considered here, this expression is linear in $\zeta$. In order to factor out the effects of transmission losses and finite detection efficiency at the intermediary site, we will compute the rescaled quantity
\begin{equation}
E  = \frac{2}{\zeta} \Tr (\mbox{${\hat{\varrho}'}$}^{(\pm)}_{\text{out}}) E_F \left( \frac{\mbox{${\hat{\varrho}'}$}^{(\pm)}_{\text{out}} }{
\Tr (\mbox{${\hat{\varrho}'}$}^{(\pm)}_{\text{out}})}\right)
\end{equation}
which is a function of the beam-splitter transmission $T$ and the photon statistics $\{ p_m\}$. To optimize the produced entanglement, for each set of the parameters characterizing the photon statistics we carried out maximization over the transmission $T$. The results are depicted in Figs.~\ref{fig:figure3}(a) and \ref{fig:figure3}(b), and the optimal values of $T$ are shown in Figs.~\ref{fig:figure3}(c) and \ref{fig:figure3}(d), correspondingly. It is seen that although the threshold condition given in Eq.~(\ref{Eq:EntanglementGenIneq}) is relatively relaxed, a substantial amount of entanglement is produced only for a small region of parameters close to the ideal case of perfect single-photon sources. In this region, the difference between the two models of the photon statistics becomes minor. For practical photon sources based on heralded parametric down-conversion, the realistic values of $r$ would be in the range $10^{-2}$ -- $10^{-1}$. We considered here a broader range of $r$ to characterize performance of other sources featuring a large multiphoton probability distributed over a ``tail'' extending to high photon numbers.

In order to reach rates calculated above in processing and utilizing produced entanglement, one needs to be able to perform a broad range of local operations on quantum memories storing noisy entangled states.
An alternative question is whether it would be possible to verify directly the presence of entanglement via, e.g., violation of Bell's inequalities. When measurements are based on photon-counting detectors, the one-photon state $\ket{\psi_\pm}$ defined in Eq.~(\ref{Eq:psipm}) is not sufficient on its own to demonstrate correlations incompatible with local hidden variable theories \cite{SinglePhotonNonlocality}. However, if in addition a local phase reference is available, one can implement noncommuting measurements whose results violate Bell's inequalities. We will consider here a scheme that relies on applying a phase-space displacement by mixing the field with a strong coherent field on an unbalanced beam splitter and counting photons in the transmitted beam \cite{BanaWodPRL1999}. For binary detectors, in the asymptotic limit of unit beam-splitter transmission, this realizes a projection on a coherent state whose amplitude is given by the displacement introduced by the reference field. When the measurement is performed on fields released from the memories, the probability of a joint no-count event is therefore given by
\begin{equation}
\label{Eq:Qalphabeta}
Q(\alpha,\beta) = \frac{1}{\Tr (\hat{\varrho}^{(\pm)}_{\text{out}})} \bra{\alpha_{a_1} \beta_{a_2}} \hat{\varrho}^{(\pm)}_{\text{out}}
\ket{\alpha_{a_1} \beta_{a_2}}
\end{equation}
and has the interpretation, up to a multiplicative constant, of the two-mode $Q$ function for the normalized state $\hat{\varrho}^{(\pm)}_{\text{out}}$. The marginal no-count probabilities for individual detectors are given by an expression
\begin{equation}
\label{Eq:Qalpha}
Q_{a_1} (\alpha) = \frac{1}{\Tr (\hat{\varrho}^{(\pm)}_{\text{out}})} \bra{\alpha_{a_1} } \Tr_{a_2}( \hat{\varrho}^{(\pm)}_{\text{out}} )
\ket{\alpha_{a_1} }
\end{equation}
for mode $a_1$ and an analogous expression for mode $a_2$.

If two alternative coherent displacements $\alpha$ or $\alpha'$ are applied to mode $a_1$ and $\beta$ or $\beta'$ to mode $a_2$, measurements of joint and marginal probabilities can be used to evaluate the Clauser-Horne (CH) combination \cite{ClauHorne74},
\begin{multline}
{\cal CH} = Q(\alpha,\beta) + Q(\alpha', \beta) + Q(\alpha, \beta') - Q(\alpha',\beta') \\
- Q_{a_1}(\alpha) - Q_{a_2}(\beta)
\end{multline}
which for local hidden variable theories is bounded between
\begin{equation}
-1 \le {\cal CH} \le 0.
\end{equation}
In Fig.~\ref{fig:figure4} we depict the minimum value of the combination obtained by optimization over displacements $\alpha, \alpha', \beta$, and $\beta'$ for the two models of photon statistics introduced in Sec.~\ref{Sec:Sources}. The output density matrix $\hat{\varrho}^{(\pm)}_{\text{out}}$ has been taken in the limit $\zeta \ll 1$ and $T\ll 1$. The photon statistics $p_m$ based on the down-conversion model has been truncated at $m=3$, which reduces the trace of the input density matrix $\hat{\varrho}_{\text{in}}$ by less than $0.1\%$ within the relevant parameter region. We verified that including the photon statistics up to $m=6$ for the optimal displacements does not noticeably change the value of the CH combination.

It is seen that a violation of Bell's inequalities places stringent requirements on the photon statistics, much stricter than the generation of entanglement in the zero-one photon sector. Although the differences between the two-photon model and the down-conversion model for the photon statistics are rather minor, the two-photon model gives a slightly larger region over which a violation of Bell's inequality can be observed if the parameters are identified as $\epsilon = 2r$. It is also worth noting that because the phase-space displacement is a linear operation, the parameter $\eta$ characterizing the photon statistics can also include the nonunit efficiency of detectors used to implement the coherent-state projections described in Eqs.~(\ref{Eq:Qalphabeta}) and (\ref{Eq:Qalpha}).

\section{Two-photon scheme}
\label{Sec:FourMemories}

In order to prepare a pair of photons entangled with respect to a modal degree of freedom such as polarization, one could repeat the scheme described in Sec.~\ref{Sec:TwoMemories} twice and postselect the output on the presence of one excitation at each node. In this section we present and analyze a two-photon scheme that, starting from four photons, directly prepares two-photon entanglement without resorting to postselection. The scheme is a linear optics analog of the proposal based on generating twin excitations in atomic ensembles \cite{ChenZhao07,ZhaoChen07}. Its principle of operation exploits two-photon interference, thus avoiding the need for interferometric stability between the nodes.

\begin{figure}
\centering
\includegraphics[width=3in]{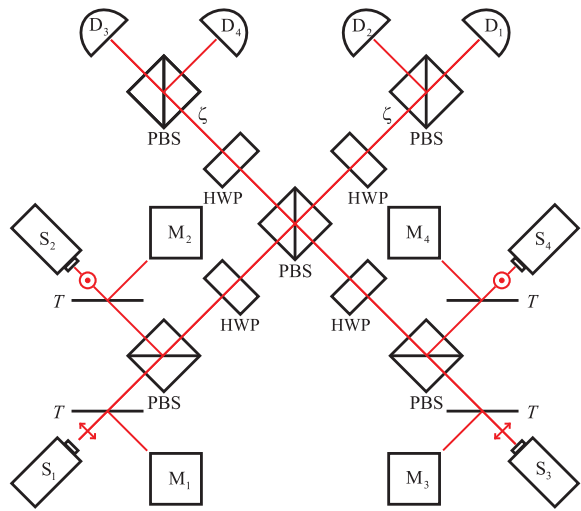}
\caption{(Color online) A scheme to generate a two-photon polarization-entangled state. Two horizontally polarized photons emitted by sources $S_1$ and $S_3$ and two vertically polarized photons emitted by sources $S_2$ and $S_4$ are partly reflected using beam splitters $T$ to quantum memories $M_1, \ldots , M_4$. The transmitted beams are combined and interfered using polarizing beam splitters (PBS) and half-wave plates (HWP), with outputs monitored by heralding detectors $D_1, \ldots , D_4$.}
\label{fig:figure5}
\end{figure}

The scheme is shown in Fig.~\ref{fig:figure5}.  Each node comprises two modes described by annihilation operators $\hat{a}_1$ and $\hat{a}_2$ for node $A$ and $\hat{a}_3$ and $\hat{a}_4$ for node $B$. It will be convenient to think of the odd-numbered modes as polarized horizontally and even-numbered modes as polarized vertically. The photons emitted by sources $S_1, \ldots, S_4$ are transmitted through beam splitters characterized by power transmission $T$, with the reflected beams stored in quantum memories $M_1, \ldots, M_4$. The pairs of modes at each node are combined into a single path using polarizing beam splitters and sent to the intermediary site where two-photon interference is realized. First, the polarizations of the two beams are rotated by $45^\circ$ using half-wave plates which realize the transformation
\begin{equation}
\hat{a}_1 \rightarrow \frac{1}{\sqrt{2}}(\hat{a}_1 + \hat{a}_2), \quad
\hat{a}_2 \rightarrow \frac{1}{\sqrt{2}}(\hat{a}_1 - \hat{a}_2)
\end{equation}
and analogously for the modes $\hat{a}_3$ and $\hat{a}_4$. Then the polarization components are recombined in the rectilinear basis on a polarizing beam splitter which transmits horizontal polarization and reflects vertical polarization. The output beams are transmitted through half-wave plates rotating polarization by $45^\circ$ and separated into horizontal and vertical components, which are monitored by detectors $D_1, \ldots, D_4$. This linear optics network implements the following transformation of the modes $\hat{a}_1 , \ldots , \hat{a}_4$:
\begin{eqnarray}
\mathbf{U}=\frac{1}{2}\left(
\begin{array}{cccc}
1 & 1 & 1 & -1 \\
1 & 1 & -1 & 1 \\
1 & -1 & 1 & 1 \\
-1 & 1 & 1 & 1
\end{array}\right).
\end{eqnarray}
In the case of perfect single-photon sources, entanglement is generated between single-photon subspaces of pairs of modes $\hat{a}_1, \hat{a}_2$ and $\hat{a}_3, \hat{a}_4$, encoding polarization qubits that are stored in quantum memories $M_1, M_2$ and $M_3, M_4$.
It is convenient to denote the basis states for the two qubits as
\begin{align}
\label{Eq:qubitbases}
\ket[A]{h} & = \ket{1_{a_1} 0_{a_2}}, & \ket[B]{h} & = \ket{1_{a_3} 0_{a_4}} \nonumber \\
\ket[A]{v} & = \ket{0_{a_1} 1_{a_2}}, & \ket[B]{v} & = \ket{0_{a_3} 1_{a_4}}
\end{align}
and define the projection operators
\begin{equation}
\hat{\Pi}_{\nu} = \proj[\nu]{h} + \proj[\nu]{v}, \quad \nu=A, B.
\label{eq:projection2ph}
\end{equation}
We retain events when exactly one of the detectors $D_1, D_2$ and one of the detectors $D_3, D_4$ click. If the single-photon sources are ideal, sequences ${\bf k} = (k_1, k_2, k_3, k_4) = (1,0,1,0)$ and ${\bf k} = (0,1,0,1)$ generate a state
\begin{multline}
{\hat{\varrho}^{({\mathbf k})}}_{\text{out}} = {\textstyle\frac{1}{2}} \zeta^2 T^2 \{ (1-T)^2 \proj{\Phi_+}  \\
 + {\textstyle\frac{1}{2}} T(1-T) [\hat{\Pi}_{A} \otimes \proj{0_{a_3} 0_{a_4}} \\ + \proj{0_{a_1} 0_{a_2}} \otimes \hat{\Pi}_{B}]
 + T^2 \proj{\text{vac}} \},
\end{multline}
where
\begin{align}
\ket{\Phi_+} & = \frac{1}{\sqrt{2}} (\hat{a}_2\hat{a}_4 + \hat{a}_1\hat{a}_3)\ket{1_{a_1} 1_{a_2} 1_{a_3} 1_{a_4}}\nonumber \\
& = \frac{1}{\sqrt{2}} ( \ket[A]{h}\ket[B]{h} + \ket[A]{v}\ket[B]{v})
\end{align}
is the maximally entangled two-photon state.
The remaining two combinations of clicks, i.e., ${\bf k} = (1,0,0,1)$ and ${\bf k} = (0,1,1,0)$, generate an analogous state with $\ket{\Phi_+}$ replaced by
\begin{equation}
\ket{\Psi_+} = \frac{1}{\sqrt{2}} ( \ket[A]{h}\ket[B]{v} + \ket[A]{v}\ket[B]{h} ).
\end{equation}
Analogously to the one-photon scheme, this state can be converted into $\ket{\Phi_+}$ by a local unitary transformation and we can restrict our attention to only one type of events.

\section{Two-photon entanglement}
\label{TwoPhotonEntanglement}

Let us now characterize entanglement generated using the scheme described in the preceding section. We will be primarily interested in the subspace spanned by the tensor products $\ket[A]{h}\ket[B]{h}$, $\ket[A]{h}\ket[B]{v}$, $\ket[A]{v}\ket[B]{h}$, and $\ket[A]{v}\ket[B]{v}$ of the qubit states defined in Eq.~(\ref{Eq:qubitbases}). The un-normalized two-photon state in this subspace can be written as
\begin{align}
\mbox{${\hat{\varrho}'}$}^{({\bf k})}_{\text{out}}  & = ( \hat{\Pi}_{A} \otimes \hat{\Pi}_{B}) \hat{\varrho}^{({\bf k})}_{\text{out}}
( \hat{\Pi}_{A} \otimes \hat{\Pi}_{B}) \nonumber \\
& = \begin{pmatrix} \varrho_{hh} & 0 & 0 & c \\ 0 & \varrho_{hv} & 0 & 0 \\ 0 & 0 & \varrho_{vh} & 0 \\ c^\ast & 0 & 0 & \varrho_{vv}
\end{pmatrix},
\label{eq:rhoprim2ph}
\end{align}
where the only nonzero elements are given explicitly in the limit $\zeta\ll 1$ and  $T\ll 1$ by
\begin{subequations}
\begin{align}
{\varrho}_{hh} = {\varrho}_{vv} & = {\textstyle\frac{1}{4}} \zeta^2T^2 (p_1^4+p_0p_1^2p_2+4p_0^2p_2^2+3p_0^2p_1p_3 ) \\
{\varrho}_{hv} = {\varrho}_{vh} & = {\textstyle\frac{1}{4}}  \zeta^2T^2 (5p_0p_1^2p_2+3p_0^2p_1p_3 )\\
c & = {\textstyle\frac{1}{4}} \zeta^2T^2  p_1^4
\end{align}
\end{subequations}
The state $\mbox{${\hat{\varrho}'}$}^{({\bf k})}_{\text{out}}$ is entangled if and only if the PPT criterion is violated, $|c|^2 > \varrho_{hv} \varrho_{vh}$, which
expressed in terms of the photon statistics takes the form $p_1 ^3 > 5 p_0 p_1 p_2 + 3 p_0^2 p_3$. When $p_3=0$, this inequality reduces to
\begin{equation}
\label{Eq:Threshold2Photon}
p_1^2 > 5 p_0 p_2,
\end{equation}
which is weaker than the criterion for the two-photon scheme given in Eq.~(\ref{Eq:EntanglementGenIneq}), but still leaves a gap compared to the classicality condition derived in Eq.~(\ref{Eq:CoherentMixtures}).

\begin{figure*}
\centering
\includegraphics[width=14cm]{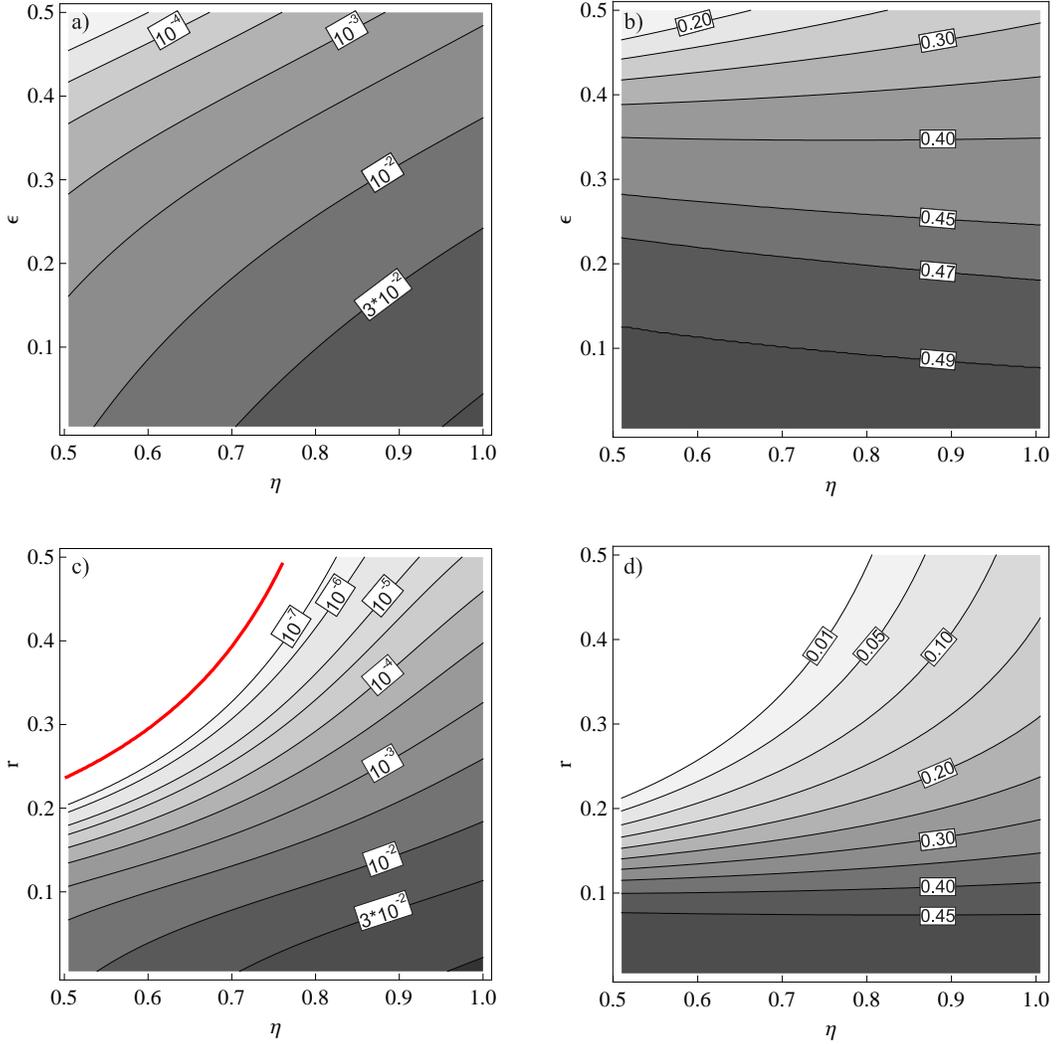}
\caption{The effective entanglement of formation defined in Eq.~(\protect\ref{Eq:EffEF2photon}) for the two-photon scheme maximized in the regime $\zeta \ll 1$ for (a) the double-emission model and (b) the down-conversion model of the photon statistics over the beam-splitter transmission $T$, with the optimal values of $T$ shown respectively in (c) and (d).}
\label{fig:figure6}
\end{figure*}

Analogously to the one-photon scheme, we quantify the amount of entanglement produced in the qubit sector using the rescaled entanglement of formation:
\begin{equation}
\label{Eq:EffEF2photon}
E  = \frac{4}{\zeta^2} \Tr (\mbox{${\hat{\varrho}'}$}^{({\bf k})}_{\text{out}}) E_F \left( \frac{\mbox{${\hat{\varrho}'}$}^{({\bf k})}_{\text{out}} }{\Tr (\mbox{${\hat{\varrho}'}$}^{({\bf k})}_{\text{out}})}\right).
\end{equation}
In the front multiplicative factor, the numerator $4$ stems from four relevant combinations of detector counts, while the denominator $\zeta^2$ is a consequence of the quadratic scaling of the scheme with the efficiency of heralding detectors, as in the present case two photons need to reach the intermediary site and be detected there.

The results of numerical optimization of the effective entanglement of formation $E$ defined in Eq.~(\ref{Eq:EffEF2photon}) for the two models of photon statistics over the beam-splitter transmission $T$ in the limit $\zeta \ll 1$ are shown in Fig.~\ref{fig:figure6}. For nearly optimal sources the effective entanglement is lower compared to the one-photon scheme. This is qualitatively understood, as in the current case twice as many photons need to be routed correctly to heralding detectors through the linear optics circuit and the qubit prepared at each node is encoded in two orthogonally polarized states of a single photon rather than a pair of zero- and one-photon Fock states. It is worth noting that the region of parameters where entanglement can be generated is larger then in Fig.~\ref{fig:figure4}, consistent with the weaker threshold condition derived in Eq.~(\ref{Eq:Threshold2Photon}).

Two-photon entanglement can be used to test Bell's inequalities with photon-counting detectors, without auxiliary reference fields. We will consider here the standard CHSH inequality \cite{CHSH69} for correlations between photon polarizations when the pairs of modes carrying the qubits are separated on polarizing beam splitters and detected with two detectors at each node. In the present case, when imperfect photon sources are used, care needs to be taken to correctly include multiphoton terms present in the conditional states $\hat{\varrho}^{({\bf k})}_{\text{out}}$.

\begin{figure*}
\centering
\includegraphics[width=14cm]{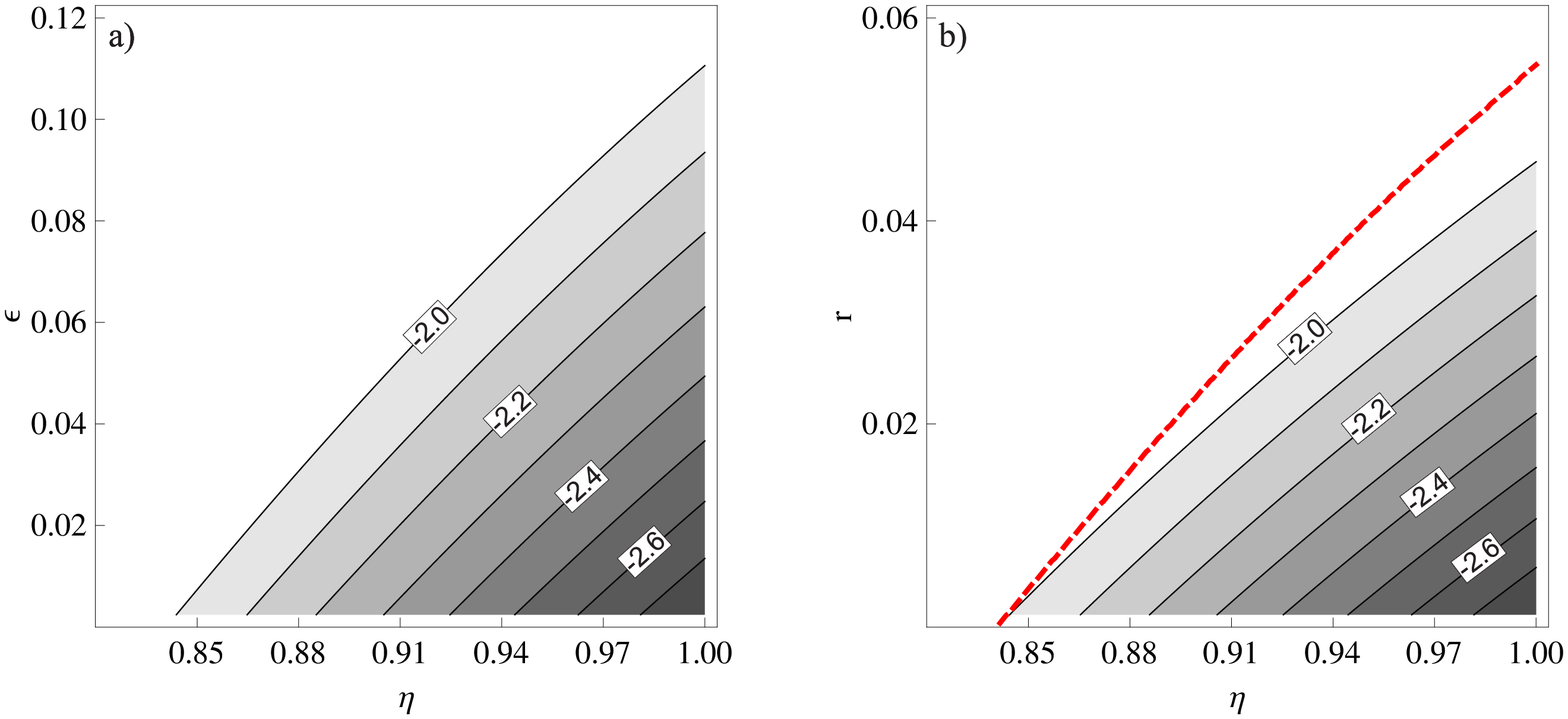}
\caption{The Clauser-Horne-Shimony-Holt combination ${\cal CHSH}$ for the two-photon polarization-entangled state generated using the scheme shown Fig.~\ref{fig:figure5} in the regime $T \ll 1$ and $\zeta \ll 1$ assuming (a) the two-photon model and (b) the down-conversion model for the photon statistics. The dashed line in panel (b) corresponds to the contour ${\cal CHSH} = -2$ in panel (a).}
\label{fig:figure8}
\end{figure*}

Rotating the polarization basis of the qubit $A$ by an angle $\theta_A$ with the help of a half-wave plate is described by a linear transformation of the annihilation operators,
\begin{align}
\hat{a}_1 & \rightarrow \hat{a}_1 \cos{\theta_A} + \hat{a}_2 \sin{\theta_A} \nonumber \\
\hat{a}_2 & \rightarrow \hat{a}_1 \sin{\theta_A} - \hat{a}_2 \cos{\theta_A}
\end{align}
which induces a certain unitary transformation $\hat{U}_A (\theta_A)$ for the modes $\hat{a}_1$ and $\hat{a}_2$. Similarly, rotation of the polarization basis for the modes $\hat{a}_3$ and $\hat{a}_4$ is described by a unitary $\hat{U}_B (\theta_B)$. The polarization-rotated four-mode state after normalization reads
\begin{equation}
\hat{\varrho}(\theta_A, \theta_B) = \frac{[\hat{U}_A(\theta_A) \otimes \hat{U}_B(\theta_B)]
\hat{\varrho}^{({\bf k})}_{\text{out}} [\hat{U}_A^\dagger(\theta_A) \otimes \hat{U}_B^\dagger (\theta_B)]}{\Tr \hat{\varrho}^{({\bf k})}_{\text{out}}}.
\end{equation}
We will assume that the output beams are monitored using binary detectors. Events when neither or both detectors clicked at one node are considered as inconclusive and carry no contribution to polarization correlation functions, but they are included in the overall normalization in order to avoid the detection loophole. The probability of a coincidence between detectors monitoring rotated beams $\hat{a}_1$ and $\hat{a}_3$ is given by
\begin{multline}
P_{13}(\theta_A, \theta_B) \\ = \sum_{k,l=1}^{\infty} \bra{k_{a_1} 0_{a_2} l_{a_3} 0_{a_4}} \hat{\varrho}(\theta_A, \theta_B)
\ket{k_{a_1} 0_{a_2} l_{a_3} 0_{a_4}},
\end{multline}
and analogously for the remaining three combinations of coincidences between the nodes described by probabilities $P_{14}(\theta_A, \theta_B)$,
$P_{23}(\theta_A, \theta_B)$, and $P_{24}(\theta_A, \theta_B)$. The polarization correlation function for given settings of polarizing beam splitters $\theta_A,\theta_B$ is expressed in terms of these probabilities as
\begin{multline}
J(\theta_A,\theta_B) = P_{13}(\theta_A, \theta_B) - P_{14}(\theta_A, \theta_B) \\ - P_{23}(\theta_A, \theta_B)  +P_{24}(\theta_A, \theta_B).
\end{multline}
The CHSH combination,
\begin{equation}
\mathcal{CHSH}=J(\theta_A,\theta_B)+J(\theta_A',\theta_B)+J(\theta_A,\theta_B')-J(\theta_A',\theta_B'),
\end{equation}
satisfies for local hidden variable theories the inequality
\begin{equation}
-2 \le \mathcal{CHSH} \le 2.
\end{equation}

In Fig.~\ref{fig:figure8} we depict the CHSH combination for the standard choice of angles $\theta_A=0$, $\theta_A' = -\pi/4$, and $\theta_B'=-\theta_B = 3\pi/8$ in the regime $T\ll 1$ and $\zeta \ll 1$, assuming the double-emission model (a) and the down-conversion model (b) of the photon statistics. In the latter case, up to $m=4$ photons have been taken into account in calculations. We verified that the results do not change noticeably within the resolution of the graphs if numerical optimization over the angles $\theta_A$, $\theta_A'$, $\theta_B$, $\theta_B'$ is performed, with the down-conversion model truncated at $m=3$. The regions where a significant violation of Bell's inequalities is possible are similar to the one-photon case, although the difference between the two models for the photon statistics is now more pronounced. This can be attributed to the deleterious effects of multiphoton terms in the input photon statistics which generate double counts at one node, thus lowering the value of the correlation function $J(\theta_A,\theta_B)$.

\section{Conclusions}
\label{Sec:Conclusions}

We studied the performance of elementary linear-optics schemes for entanglement distribution based on imperfect single-photon sources, linear optics, and heralding detectors. The underlying principle of nonlocal photon subtraction permits preparation of one-photon entanglement, where a single photon is prepared in a delocalized superposition state, and two-photon entanglement, where two photons located at different nodes are entangled in a modal degree of freedom such as polarization. Two models of photon statistics describing imperfect photon sources were considered: the first one assumes occasional double emission, and the second one describes heralded sources based on spontaneous parametric down-conversion, with a relatively long multiphoton ``tail''. Other types of photon sources can be expected to exhibit statistics that lie between these two extreme models.

We analyzed sensitivity  to photon source imperfections of entanglement generated between photon-number qubits in the one-photon scheme and polarization qubits in the two-photon scheme. Although nonseparable states are produced for a relatively broad range of parameters, a substantial amount of entanglement is obtained only for inputs close to ideal single photons. This analysis assumed implicitly that a wide range of operations can be implemented locally to process and distill noisy entanglement created in respective qubit subspaces. Distilled maximally entangled states between adjacent nodes could be used as a resource for standard entanglement swapping operations to extend the range of entanglement. In this strategy, the entanglement measure analyzed in this paper can be used as an indicator of overheads resulting from the use of imperfect resources that should be included in the standard analysis of the performance of quantum repeaters \cite{SangSimon07,SimondeRied07}. More generally,
an important task is to develop feasible and efficient methods for extending the range of entanglement in realistic quantum repeater architectures and to analyze the effects of imperfections beyond an elementary link \cite{1404.7183}.

A complementary question is whether the generated bipartite state can be used ``as is'' to test Bell's inequalities. For the one-photon scheme, auxiliary coherent reference beams are needed to implement noncommuting measurements based on photon counting, while in the two-photon case polarization measurements are sufficient. A construction of an elementary link for entanglement distribution with high-fidelity quantum memories should in principle permit a loophole-free violation of Bell's inequalities in a regime when only a lossy optical channel is available between the nodes. If this is the primary objective, one could consider generation of nonmaximally entangled states, which may be more robust for an imperfect read out of quantum memories \cite{EberhardPRA1992}.

We expect that this detailed study on the statistics of sources will be of great practical benefit with an increasing number of experimental systems studied to develop quantum repeater links, as well as from a more fundamental perspective, to perform loophole-free tests of Bell's inequalities.

\section*{Acknowledgements}

We wish to acknowledge insightful discussions with Rafa{\l} Demkowicz-Dobrza\'{n}ski and Norbert L\"{u}tkenhaus. This work was supported by a Foundation for Polish Science TEAM Project cofinanced by the EU European Regional Development Fund and the European Union 7th Framework Programme Project SIQS (Grant Agreement no.\ 600645).


\begin{thebibliography}{20}

\bibitem{GisiRiboRMP2002}
N. Gisin, G. Ribordy, W. Tittel, and H. Zbinden, Rev. Mod. Phys. \textbf{74}, 145 (2002).

\bibitem{ScarBechRMP2009}
V. Scarani, H. Bechmann-Pasquinucci, N. J. Cerf, M. Du\v{s}ek, N. L\"{u}tkenhaus, and M. Peev, Rev. Mod. Phys. \textbf{81}, 1301 (2009).

\bibitem{BriDurPRL1998}
H.-J. Briegel, W. D\"{u}r, J. I. Cirac, and P. Zoller, Phys. Rev. Lett. \textbf{81}, 5932--5935 (1998).

\bibitem{BennBrasPRL1993}
C. H. Bennett, G. Brassard, C. Cr\'{e}peau, R. Jozsa, A. Peres, and W. K. Wootters, Phys. Rev. Lett. \textbf{70}, 1895 (1993).

\bibitem{ZukoZeilPRL1993}
M. \.{Z}ukowski, A. Zeilinger, M. A. Horne, and A. K. Ekert, Phys. Rev. Lett. \textbf{71}, 4287--4290 (1993).

\bibitem{LukinRMP2003}
M. D. Lukin, Rev. Mod. Phys. \textbf{75}, 457--472 (2003).

\bibitem{SangSimonRMP2011}
N. Sangouard, C. Simon, H. de Riedmatten, and N. Gisin, Rev. Mod. Phys. \textbf{83}, 33--80 (2011).

\bibitem{DuanLukinNAT2001}
L.-M. Duan, M. D. Lukin, J. I. Cirac, and P. Zoller, Nature (London) \textbf{414}, 413--418 (2001).

\bibitem{JiangTay07}
L. Jiang, J. M. Taylor, and M. D. Lukin, Phys. Rev. A \textbf{76}, 012301 (2007).

\bibitem{ZhaoChen07}
B. Zhao, Z.-B. Chen, Y.-A. Chen, J. Schmiedmayer, and J.-W. Pan, Phys. Rev. Lett. \textbf{98}, 240502 (2007).

\bibitem{ChenZhao07}
Z.-B. Chen, B. Zhao, Y.-A. Chen, J. Schmiedmayer, and J.-W. Pan, Phys. Rev. A \textbf{76}, 022329 (2007).

\bibitem{SangSimon08}
N. Sangouard, C. Simon, B. Zhao, Y.-A. Chen, H. de Riedmatten, J.-W. Pan, and N. Gisin, Phys. Rev. A \textbf{77}, 062301 (2008).

\bibitem{SimondeRied07}
C. Simon, H. de Riedmatten, M. Afzelius, N. Sangouard, H. Zbinden, and N. Gisin, Phys. Rev. Lett. \textbf{98}, 190503 (2007).

\bibitem{SangSimon07}
N. Sangouard, C. Simon, J. Min\'{a}\v{r}, H. Zbinden, H. de Riedmatten, and N. Gisin, Phys. Rev. A \textbf{76}, 050301 (2007).

\bibitem{DaknaAnhut97}
M. Dakna, T. Anhut, T. Opatrn\'{y}, L. Kn\"{o}ll, and D.-G. Welsch, Phys. Rev. A \textbf{55}, 3184--3194 (1997).

\bibitem{KimPark05}
M. S. Kim, E. Park, P. L. Knight, and H. Jeong, Phys. Rev. A \textbf{71}, 043805 (2005).

\bibitem{Kim08}
M. S. Kim, J. Phys. B: At., Mol. Opt. Phys. \textbf{41}, 133001 (2008).

\bibitem{CastellettoScholten}
S. A. Castelletto and R. E. Scholten, Eur. Phys. J.: Appl. Phys. \textbf{41}, 181--194 (2008).

\bibitem{MigdallBranningPRA2002}
A. L. Migdall, D. Branning, and S. Castelletto, Phys. Rev. A \textbf{66}, 053805 (2002).

\bibitem{ShapiroWongOL2007}
J. H. Shapiro and F. N. C. Wong, Opt. Lett. \textbf{32}, 2698–-2700 (2007).

\bibitem{CollinsXiongNCOMM2013}
M. J. Collins, C. Xiong, I. H. Rey, T. D. Vo, J. He, S. Shahnia, C. Reardon, T. F. Krauss, M. J. Steel, A. S. Clark, and B. J. Eggleton, Nat. Commun. \textbf{4}, 2582 (2013).

\bibitem{ShieldsNPHOT2007}
A. J. Shields, Nature Photon. \textbf{1}, 215--223 (2007).

\bibitem{EisamanFanRSI2011}
M. D. Eisaman, J. Fan, A. Migdall, and S. V. Polyakov, Rev. Sci. Instrum. \textbf{82}, 071101 (2011).

\bibitem{HillWoot97}
S. Hill and W. K. Wootters, Phys. Rev. Lett. \textbf{78}, 5022--5025 (1997).

\bibitem{Woot98}
W. K. Wootters, Phys. Rev. Lett. \textbf{80}, 2245--2248 (1998).

\bibitem{HoroHoroRMP2009}
R. Horodecki, P. Horodecki, M. Horodecki, and K. Horodecki, Rev. Mod. Phys. \textbf{81}, 865--942 (2009).

\bibitem{MandelJOSA1977}
L. Mandel, J. Opt. Soc. Am. \textbf{67}, 1101--1104 (1977).

\bibitem{Bell87}
J. S. Bell, \textit{Speakable and Unspeakable in Quantum Mechanics} (Cambridge University Press, Cambridge, UK, 1987).

\bibitem{ClauHorne74}
J. F. Clauser and M. A. Horne, Phys. Rev. D \textbf{10}, 526--535 (1974).

\bibitem{BanaWodPRL1999}
K. Banaszek and K. W\'{o}dkiewicz, Phys. Rev. Lett. \textbf{82}, 2009--2013 (1999).

\bibitem{WallenVog96}
S. Wallentowitz and W. Vogel, Phys. Rev. A \textbf{53}, 4528--4533 (1996).

\bibitem{BanaWod96}
K. Banaszek and K. W\'{o}dkiewicz, Phys. Rev. Lett. \textbf{76}, 4344--4347 (1996).

\bibitem{CHSH69}
J. F. Clauser, M. A. Horne, A. Shimony, and R. A. Holt, Phys. Rev. Lett. \textbf{23}, 880--884 (1969).

\bibitem{ChuangLeungPRA1997}
I. L. Chuang, D. W. Leung, and Y. Yamamoto, Phys. Rev. A \textbf{56}, 1114--1125 (1997).

\bibitem{WasiBanaPRA2007}
W. Wasilewski and K. Banaszek, Phys. Rev. A \textbf{75}, 042316 (2007).

\bibitem{SekatskiJPB2012}
P. Sekatski, N. Sangouard, F.Bussi\`{e}res, C. Clausen, N. Gisin, and H. Zbinden, J. Phys. B: At., Mol. Opt. Phys. \textbf{45}, 124016 (2012).

\bibitem{Peres96}
A. Peres, Phys. Rev. Lett. \textbf{77}, 1413--1415 (1996).

\bibitem{Horodeckis96}
M. Horodecki, P. Horodecki, and R. Horodecki, Phys. Lett. A \textbf{223}, 1--8 (1996).

\bibitem{SinglePhotonNonlocality}
L. Hardy, Phys. Rev. Lett. \textbf{73}, 2279--2283 (1994); L. Vaidman, \emph{ibid}. \textbf{75}, 2063 (1995); D. M. Greenberger, M. A. Horne, and A. Zeilinger, \emph{ibid}. \textbf{75}, 2064 (1995).

\bibitem{1404.7183}
Z. Dutton, C. A. Fuchs, S. Guha,  and H. Krovi, arXiv:1404.7183.

\bibitem{EberhardPRA1992}
P. H. Eberhard, Phys. Rev. A \textbf{47}, R747--R750 (1993).

\end{thebibliography}
\end{document}